\documentstyle{article}
\begin{document}
\large

\centerline{\Large \bf Rotation curves of spiral galaxies: A general-relativistic model}

\vskip 0.9cm

\begin{center}
\centerline {\large P. S. Negi \footnote{e-mail: negi@upso.ernet.in} } 

{\large Department of Physics, Kumaun University, Nainital-263 002, India}

\vspace {1.5cm}
{\large \bf Abstract}
\end{center}

\vspace {0.8cm}

\noindent 
Spiral  galaxies are considered as  static  and   spherically symmetric 
Dark Matter Configurations (DMC) [described by Tolman's type VII solution with vanishing
surface density] of size $a \sim 10 $ kpc in which 
$\sim  10^{11}$ Sun-like  stars [non-zero rest-mass particles (NZRPs)] move  along 
appropriate trajectories. Using general relativity (GR), 
we show that a mass of dark matter about  127  -  212 $\times 10^{11}  M_{\odot}$ is 
required inside the sphere of size $a \sim 10 $ kpc for agreement with the observed typical orbital velocity
ranging from 150 to 250 \, km\, sec$^{-1}$.
That is, on the average, the actual total mass accumulated in 
spiral galaxies is 
$M \cong (2.5-4.2) \times 10^{13} M_{\odot}$. Thus, 
even though the observed orbital velocities $v$ satisfy the condition ($v \ll c$),
GR may have important consequences.
 
In our model, it is possible to obtain flat, slightly rising, 
and even declining rotation curves,  which 
in  fact  represent  the  loci  of   various   NZRPs   originating 
simultaneously from near-central to outer  regions  of  the 
galaxy. The model self-evidently explains the reason why most of the spirals
show almost flat or slightly rising rotation curves and very few show  declining curves. The NZRPs follow 
trajectories coming very  close 
to the centre and are finally trapped in circles  of  minimum  radii $\sim 0.03a - 0.005a$ with 
rotation velocities reaching $ \sim 6544 - 8445 $ km\, sec $^{-1}$. This
is also true for NZRPs emitted in near-central regions, which get trapped even deeper. The
velocity values are consistent with observations of central
regions of various galaxies. This scenario may also lead to conditions suitable
for evolution of relativistic supermassive stars and supermassive black holes
at the centres of various galaxies.

Based on this  study,  we obtain  the density
parameter $\Omega  \sim 0.127 - 0.212 $,
leading  to an open model of the Universe with an age of $\sim 
16.8 - 17.6 $ Gyr [for the Hubble constant $H = 50$ km\, sec$^{-1}$\, Mpc$^{-1}$],  which  is 
significantly higher than the presently estimated age of globular  clusters $\sim 13 - 15 $ Gyr.

\vspace {0.8cm}
   
\vspace {5.0cm}
\pagebreak

\noindent {\large \bf 1.\,\,\, Introduction}

\vspace {0.5cm}
\noindent
For the recent three decades, a  large  amount  of  observational 
data on nearly flat and slightly rising  rotation  curves 
of spiral galaxies have pointed out that a lot  of  dark 
mass dominates in  these  galaxies (see, e.g., Rubin [1]).
An  explanation  of  these 
phenomena on the basis of the Newtonian gravitation  theory  (NGT) requires
dark matter dominating in the halo region of various 
galaxies. But the amount of the dark matter which prevails  among 
such galaxies cannot be made certain by using NGT-based models. 
However, some spiral galaxies with  sharply  declining  rotation 
curves are also observed [2, 3].  Casertano and van Gorkom [3] 
interpreted the decrease in rotation velocity as an indication  of 
a large ratio of luminous to dark matter in the  luminous  regions 
of these galaxies and  thus  weakening  of the  well-known 
``conspiracy" between luminous and dark 
matter [4, 5]. Thus,
in order to explain simultaneously all three types of rotation curves, the models based upon NGT  require individual assumptions  
like dominance of dark or luminous
matter in the halo or in the luminous region of an individual galaxy, depending upon
the nature of the rotation curve. So, no ``common ground'', {\em in general}, is available within NGT 
in explaining all
three types of rotation curves 
observed in various spirals.  The  models  based on
the Modified Newtonian Theory (MONT) [6] (see, e.g., [7]) 
[of course, this theory requires no dark 
matter at all] would also require individual assumptions for individual galaxies,
 as NGT does,  if 
the observational evidence of all three types of rotation curves 
is included.
\medskip

        Beside this problem, a number  of  observations 
regarding the central regions of various galaxies (including spiral 
galaxies) are  now  available,  which  indicate  the  existence  of 
extremely high velocity ($\sim 10^4 $ \,km \,sec $^{-1}$) of gas  clouds
and  stars around the galactic centre [8--14]. The most widely  believed
explanation  of this phenomenon is that supermassive black holes (SBHs) of masses
$\sim 10^6 - 10^9 M_{\odot}$ are present at the centres  of  various  galaxies. 
However, Maoz [15] emphasized that the presence of  such 
SBHs will be proved only when  various  other 
options (such as, clusters of brown dwarfs, or low mass  solar 
type  stars)  could  be  completely  ruled  out.  The 
presence  of such SBHs at the centres of various galaxies
would also  require a general-relativistic treatment,  instead  of the conventional  
Newtonian  formalism. Based upon this argument and the belief  that  general
relativity theory (GR) might explain  phenomena  at  large  scale,  various 
authors have used GR to explain the nearly flat and  slightly  rising 
rotation  curves  of  spiral  galaxies [16--18]. Among various such models, the authors have generally
considered static, spherically symmetric  mass  distribution, 
with a core  specifying luminous  matter  and  the  halo 
governed by a $(1/r^2)$ density distribution [so that the mass in the halo region turns out to be  proportional 
to  the  radial  distance] with the equation of state $P = \gamma \rho$ (where  $P$  and  $\rho$ are the  pressure  and 
the energy-density, respectively, and $\gamma$ is a constant).  To  terminate  the $ (1/r^2) $  density 
distribution at a  finite  radius  (since this 
distribution in GR corresponds to a non-terminating solution), an 
envelope governing some other  density  distribution  representing 
the luminous  matter  is  matched  to  the  galactic halo. 
Apparently, these  models  automatically  give  rising 
rotation curves in the halo region of the  galaxies,  which  
turn out to be nearly flat in the Newtonian limit (i.e., for the speed of light $c$  approaching 
infinity), because the expression for the rotation velocity of a 
particle (i.e, star) is worked out by  assuming  {\em arbitrarily} 
that the all particles move in circular  orbits  (see, e.g., [19]). The slope of these curves 
depends on the density distribution considered for the halo region, as is  discussed in detail in Sec.3 of 
the present paper.
\medskip

        Thus, like NGT and MONT, various GR-based models as is
discussed above, cannot provide a  suitable  explanation, 
{\em in general}, for all three  types  of  rotation  curves  observed 
among various spiral galaxies. Furthermore, these models do not provide
any reason(s) behind high-velocity gas  clouds  and 
stars observed around the central region of various galaxies.
\medskip

Therefore, at present, the main problems  associated  with  spiral
galaxies can be summarized as follows: (i) to construct a  suitable  model 
of spiral galaxies which could possibly explain, in general, all  three  types 
of observed rotation curves,  and  (ii)  to give a reasonable explanation to high-velocity gas clouds  and 
stars observed near the central region of various galaxies, and (iii) if 
SBHs are, in fact, present at the centres of various galaxies, to
trace out an actual scenario which could lead to their formation.
\medskip

        Taking all these points into account and realizing that 
GR is  a  theory  of  geodesics  which represent  the  intrinsic 
properties  of  the  space-time  geometry,  produced  by  a   mass 
distribution, we take, as a basis of our study, some  well-known  
observational data  and   avoid 
arbitrary assumptions, such as ``all the particles move  in  
circular orbits''. We specify a spiral galaxy as a static, 
spherically  symmetric  dark  matter  configuration  (DMC) in 
hydrostatic equilibrium,  in  which trapped luminous nonzero
rest-mass particles (NZRPs), such  as  stars  and 
gas  clouds, move  along  their  appropriate  trajectories 
provided by the space-time curvature due to  DMC.  In  fact,  our 
basic assumption about the status of  luminous  matter  in  spiral 
galaxies as test-particles (unable to alter the  space-time 
curvature of the whole galaxy specified as a DMC) turns out 
to be a consequence of the present study (this is what the GR demands) 
as is discussed later.
\medskip

    [Notice that here we are only proposing  an alternative model of
spiral galaxies based on broad observational evidences.
We now do not consider, the problems related to
protogalaxies, evolution of galaxies, etc.  What  we  
actually discuss is the location of  NZRPs  inside the  galaxy, 
their initial velocities at this location, and  their initial  direction. These  {\em initial  conditions}  of 
various NZRPs within the galaxy are worked out in  such  a  manner 
that a general explanation of all three types of rotation  curves,
as well as the presence of high-velocity gas clouds near the  central 
regions of various galaxies could be achieved.]
\medskip

        In NGT, the various galaxies are treated  merely 
Newtonian because the mean velocity of  constituent  stars  turns 
out to be around 300 km\, sec$^{-1}$, which corresponds to a  dimensionless 
gravitational potential $(\phi/c^2) \sim 10^{-6}$ (equivalent to  the  
``compaction parameter'', to be discussed later) for the whole galaxy [20].  This  result 
is based on considering a spiral galaxy as a  system  of 
self-gravitating particles [e.g.,  a  normal  spiral  galaxy  contains 
around $10^{11}$ Sun-like normal stars, each having 
a mass around $1 M_{\odot}$ and a radius of the order of $10^{6}$ km,  in 
a sphere of radius around 10 kpc]. The gravitational force due to
constituent stars and  gas  clouds is there balanced by the centrifugal force
generated  by  the appropriate rotation
of  whole system around the galactic centre. In this manner, each constituent 
of the galaxy can be regarded as moving in a circular orbit around the galactic
centre, so that the locus of various particles constitutes a spiral-like curve.
However, it should be noted that  even  for  the  planet  Mercury, if we use GR,
trajectories in the {\em exterior} field of the Sun (corresponding
to the same value of dimensionless gravitational potential $\sim 10^{-6}$ on the surface) turn out to be very different from those 
obtained in NGT [though, one 
can use higher-order  corrections  in NGT (with respect to $(\phi/c^2) \sim 10^{-6}$ and higher) to retrieve the results
of GR, but it would increase complications 
of the problem, rather than simplifying them]. Therefore, it may be argued that if we adopt a spiral 
galaxy as a DMC model 
and study the 
motion of  NZRPs using GR, the results should be quite different from those 
of NGT or of the earlier GR-based
models of spiral galaxies.  The major results
in this direction indicate
that, apart from the nature of trajectories obtained for NZRPs, the DMC models could have the 
values of the surface dimensionless  gravitational  potential $\sim 10^{-4}$, that is,
around two orders of magnitude higher than that obtained by using the NGT.
\medskip

Instead of assuming a spiral galaxy as  a  system 
of self-gravitating particles, we describe it as a DMC 
model with a radius of the order of 10 kpc [such a region contains nearly all luminous matter
present in spiral galaxies], in which around  $10^{11}$  Sun-like 
stars move along trajectories which  can  be  worked 
out by imposing suitable initial conditions. Note  that, 
in this model, all constituent stars do not necessarily move  in 
circular orbits at each radial distance from the centre, as  is 
necessary in the Newtonian formulation. The present type of model is also inspired by the results of 
various cosmological studies which indicate that more than $90\%$ of  the  mass  of  the 
Universe is contained in the dark component of unknown nature; it interacts with  the  luminous 
matter only through the gravitational interaction. We wish to accumulate 
this large amount of dark matter inside spiral galaxies as DMC models considered
in the present study. As in the Newtonian  case, 
the locus of various NZRPs situated  from  near the centre  to  the 
surface of the galaxy, comprise a spiral-like  curve,  but  their 
orbital velocities are  regarded  as part of the
{\em initial conditions}. By a suitable  choice  of 
these ``initial conditions'',  we  can  obtain models  of  spiral 
galaxies corresponding to nearly flat, slightly  rising,  or  even 
declining rotation curves.
\medskip

	For example, we  can  use  the  observed  mean  velocity  of 
constituent stars about 300 km\, sec$^{-1}$ as an {\em initial}  velocity  of  a 
star located at a particular position (radial  distance  from  the 
centre) with proper direction. 
Apparently, the dimensionless gravitational potential of  the  DMC 
will not necessarily correspond to a value $\sim 10^{-6} $ as 
in the Newtonian case. Instead, we have to  work  out  its 
value by imposing some other  observational  constraint  regarding 
these galaxies. Such a constraint could follow, e.g., 
from the typical values of observed rotation velocities
in various types of spiral galaxies, corresponding to the range 
of 150 to 250 km\, sec $^{-1}$ from type Sc to type Sa,  respectively.
Imposing these constraints and using the general expression for orbital velocity 
in GR, we obtain 
the compaction parameter, equivalent  to  the   dimensionless 
gravitational potential: $u[\equiv$ total mass  to  size  ratio,  $(M/a)$,  in 
geometrized  units (see the Appendix for definitions). The range of $u$ is $u \simeq (1.27 - 2.12) \times 
10^{-4})$. Note that the gravitational
potential obtained in this manner using GR turns out to be
about two orders of magnitude greater than the presently believed value and cannot 
be obtained using NGT or MONT. Thus, it  is 
apparent that even when the ratio, $(v^2/c^2)$ [$v$ is the test-particle velocity 
and $c$ the speed of light in vacuum] is small  compared  to 
unity, GR may have important consequences, in particular for 
{\em test-particle motion} (rather than {\em merely structural properties})
in the {\em interior} field of DMC-like  structures, 
as models of real galaxies.
\medskip

It follows that spiral galaxies as DMC-like structures may contain 
a dark mass which is 127 - 212  times  higher 
than the luminous mass (observed in the  form  of  stars  and  gas 
clouds in the spiral arms), independently of the  nature  of 
the rotation curve, be it flat, slightly rising or  even 
declining. (The latter is unlikely in the Newtonian theory, in which, to explain a  sharply declining rotation curve, a spiral 
galaxy must  have 
luminous matter dominance in the luminous region.) Thus  the  DMC  model considered in the present study not only justifies 
our basic assumption about spiral  galaxies  (that the  luminous 
matter can be treated as test-particles moving  inside  the  DMC), 
but also provides a suitable general logical  explanation  of 
all observed types of rotation curves, which was lacking in Newtonian models.
\medskip

        Another interesting feature of this study, which  can  not 
be obtained in NGT,  is  that  various  trapped  NZRPs form
a spiral-like  curve, corresponding to the
trajectories (shown in Fig. 2 - 4 by dashed lines) approaching 
very close to the centre and then trapped always into a circular 
orbit of minimum radius $r_{min}$ given by Eq.(27), that
is, circular orbits in GR-models are possible only at 
a minimal radius, as opposed to any radius in NGT 
(in which the motion of every constituent of the galaxy  in 
a circular orbit is necessary to 
maintain the equilibrium). The final orbital velocities of such trapped NZRPs 
(in circular orbits) reach $\sim 10^4 $ km\, sec$^{-1}$,  
which  is 
consistent with  the  observations  regarding the central  regions  of 
various galaxies. Thus the DMC models of  spiral  galaxies 
studied here may provide  a   suitable 
alternative to the belief that there are SBHs at
the  centres  of  various  galaxies. The point is that, if the high velocity of gas 
clouds and stars are explained by the presence of SBHs, the 
rotation curves near galactic centres  should  necessarily  be 
Keplerian in nature; however, this feature is not supported  by  recent 
observations of central rotation curves of spiral galaxies which
generally show a steep nuclear rise and a high rotation velocity, common
to all massive galaxies [21]. If, however, the future observations prove the existence of SBHs  at the centres of 
various galaxies, the present scenario can provide
conditions, suitable for evolution of relativistic supermassive stars and 
SBHs [22]. Though,
we do not want to stress this point any more; this scenario
could provide an appropriate basis for future studies.
\medskip

     In Sec.2, we obtain the necessary condition for  trapping of
NZRPs at various points inside the DMC.  For  this  purpose,  we 
obtain the  trapping  angle $\psi_0$,  that  is,  the  maximum 
semi-angle of the ``cone of avoidance'' (an emitted NZRP exceeding $\psi_0$
is trapped by the DMC).
\medskip

     In Sec.3,  we obtain an expression  for the orbital
velocity of an NZRP from the equations of motion in GR.
Variation of the NZRP orbital velocity is found to be inversely 
proportional to the radial distance $r$. In other  words,  for  a given
NZRP, one always  obtains  a  declining  rotation
curve  from  near  the  centre  to  the  outer   region   of   the 
configuration. However, it should be noted  that measurement 
along such a curve for an assigned NZRP on the scale of  a 
galaxy has no practical significance since it would  require  a 
period of the order of  megayears.  Then,  what can  be  the 
reason for the  large  number  of  observational 
data on flat and slightly rising rotation curves in 
most of the spiral galaxies? This question can  be well answered
if we  assume  that  we  are at present,  in  fact, {\em watching 
the locus of various trapped  NZRPs  inside  the  DMC, 
which comprise the spiral-like curve, and measuring their  initial 
orbital velocities at the moment of their origin}.
\medskip

     To understand this scenario  more  clearly, we show in  the
next section that test-particle trapping
depends only on the {\em initial  conditions}, including the initial 
orbital velocity. And these initial conditions, in fact, allow us 
to obtain flat, rising, or even declining  orbital  velocities  of 
various  NZRPs,  originating  simultaneously  from  various  points 
inside the DMC. The {\em loci} of these NZRPs  can  be  easily  arranged 
according to the observational situation, e.g., any  specific shape of the spiral.
Furthermore, we also obtain the corresponding NZRP trajectories 
for various  initial  conditions  adopted  in  this 
paper. As is discussed later, the {\em initial} orbital velocities at various
points inside the structure depend mainly on the emission angle $\psi_i$,
which provides the reason why the majority of spirals show almost flat or
slightly rising rotation curves.
\medskip
 
     Furthermore, it is well known  that  the  population  of  the
Universe is dominated by spiral galaxies. The density parameter $ \Omega \equiv \rho_0 /\rho_{crit}$ (the ratio
of present mass density to the critical density of the Universe) $ \sim  0.002  -  0.003 $ for 
galaxies  alone, which includes a dark  mass of 3 to 5 times the mass of luminous matter [1, 23, 24]. However, if  we  assume  that 
most of the dark matter in the Universe  is more  or 
less  distributed  among  galaxies (in a manner in which spiral  galaxies  may 
contain a dark mass of 127 - 212 times the mass of luminous matter), then the density parameter will be at 
most $\Omega \sim $ 0.127  -  0.212, leading to an open model of  the  Universe  with  an  age 
estimate of about 16.8 - 17.6 Gyr (for the Hubble constant $H = 50\, {\rm km\, sec}^{-1} \, {\rm Mpc}^{-1}$ as used by Rubin [1] for observations of  spiral 
galaxies). This age estimate is significantly higher than the  age 
of globular clusters $\sim 13 - 15 $ Gyr [25, 26].

\vspace {0.5cm} 

\noindent {\large \bf 2.\,\,\, Equations of motion and test-particles trapping in static, spherically symmetric mass distributions}
          
\vspace {0.5cm}          
\noindent
   The  metric  corresponding  to  a  static, spherically 
symmetric mass distribution can be written as
\begin{equation}
ds^2  =  e^{\nu} dt^2  -  e^{\lambda} dr^2 - r^2 d\theta^2  - r^2 \sin^2 \theta d\phi^2, 
\end{equation}
where $G = c = 1$ and  $\nu$  and $\lambda$  are  functions  of  $r$  alone. The
resulting field equations for systems with isotropic  pressure  $P$ 
and energy-density $\rho$ are well known and  comprise the standard  textbook
material.
\medskip

     Defining an angle $\psi$ [as shown in Fig.1, in  which  we
use, for convenience, the radial  distance  $y (\equiv r/a)$ in  units  of 
configuration size] as  the  angle  between  the 
radial direction and tangent to the orbit $r = r(\phi)$ in the $\theta  =  (\pi/2)$
plane, we can write
\begin{equation}
\tan \psi  =  r e^{-\lambda /2} (d\phi /dr).
\end{equation}
Let $J$ be the  specific   angular   momentum   [(angular
momentum/rest-mass)]   and   $E$    be    the    specific    energy 
[(energy/rest-mass)] of a test-particle  in  this  plane, so that the 
ratio $(J/E)$ represents the ``impact parameter'' and is  a  constant 
of motion [20, 27, 28], that is
\begin{equation}
b  = (J/E).
\end{equation}
For the metric (1), the equations of motion 
are [29]
\begin{eqnarray}
\theta & = & {\rm constant} = (\pi /2) {\rm \,\,say},  \\
r^2 (d\phi /ds) & = & J  =  {\rm constant},  \\
e^{\nu} (dt/ds) & = & E  =  {\rm constant},  \\
e^{\lambda} (dr/ds)^2 & = & e^{-\nu} (E^2 - e^{\nu} [(J/r)^2 + 1]).  
\end{eqnarray}
Eqs.(3), (4), and (7) indicate that the orbit
is confined to a plane (for convenience, we take it as the equatorial
plane, $\theta = \pi /2)$.
In the metric (1), we get
\[e^{\lambda} (dr/ds)^2  = e^{\nu} (dt/ds)^2  -  r^2 (d\phi /ds)^2  -  1. \]
Applying the conditions (5) and (6), we obtain
\[e^{\lambda} (dr/d\phi)^2  =  e^{-\nu} (E^2 /J^2) r^4  -  (r^4 /J^2) - r^2  \]
and finally, using Eq.(3) for the impact parameter, we get
\begin{equation}
e^\lambda (dr/d\phi)^2  =  r^2 e^{-\nu} (r/b)^2  -  r^2 E^{-2} (r/b)^2 - r^2, 
\end{equation}
or,
\begin{equation}
(d\phi /dr)  =  (b/r) e^{-\lambda/2}/{[r^2 (e^{-\nu} - E^{-2}) - b^2]}^{1/2},
\end{equation}
which is the required equation  describing  the test-particle trajectory
 [an NZRP or ZRP (zero rest-mass)].  The  difference  in 
the ZRP and NZRP trajectories follows from the  term  ($r^2 E^{-2}$)  in 
the denominator of the right hand side of Eq.(9) [30].
\medskip

From Eqs.(2) and (8), we get
\begin{equation}
\sin \psi  =  (b/r)(e^{-\nu} - E^{-2})^{-1/2},
\end{equation}
when $\psi  =  90^{\circ}$, 
\begin{eqnarray}
b & = & b(90^{\circ})  =  r(e^{-\nu} - E^{-2})^{1/2},  
\end{eqnarray}
and $b(90^{\circ})$ attains its minimum value, $b_{min}(90^{\circ}) = \ss$, when
\begin{equation}
(d\nu /dr)  =  (2/r)(1  -  e^{\nu} E^{-2}).
\end{equation}
Now, defining an angle $\psi_0$ given by
\begin{equation}
\sin \psi_0   =  (\ss /r) (e^{-\nu} - E^{-2})^{-1/2},
\end{equation}
whose physical importance lies in the fact the when $\psi \ > \psi_0$, the 
emitted test-particles will be gravitationally trapped  inside the 
spherical  configuration.  Thus $\psi_0$  represents   the   maximum 
semi-angle of the cone at a radial  distance $r$ such  that  only 
particles emitted inside this cone will  be  able  to 
escape the mass distribution.
\medskip

     Let $u [\equiv (M/a)$, total mass to size ratio] represents the  compaction 
parameter of the spherical configuration, and, for convenience,  we will
measure the impact parameter $b$ and the radial coordinate $r$ in 
units of the configuration size $ a $, so that
\begin{equation}
\beta \equiv (b/a), {\rm and}  
\end{equation}
\begin{equation}
y \equiv (r/a).  
\end{equation}
In terms of these variables, Eqs.(9), (10), and (13) become 
\begin{equation}
(d\phi /dy) = (\beta /y)e^{-\lambda/2}/{[y^2 (e^{-\nu} - E^{-2}) - \beta^2]}^{1/2},
\end{equation}
\begin{equation}
\beta  =  y(e^{-\nu}  -  E^{-2})^{1/2} \sin \psi,
\end{equation}
\begin{equation}
\sin \psi_0  =  (2u\ss /y){(e^{-\nu} - E^{-2})}^{-1/2}.
\end{equation}

\vspace {0.5cm}  
\noindent {\large \bf 3.\,\,\, General expression for the NZRP orbital velocity inside a DMC}

\vspace {0.5cm}

The orbital velocity  $v_{\phi}$ of  a 
NZRP moving inside the DMC governed by the  metric (1) is
\begin{equation}
v_\phi  =  r(d\phi /dt),
\end{equation}
which can be written as
\begin{equation}
v_\phi   =  r(d\phi/ds)(dt/ds).
\end{equation}
Substitution of Eqs.(5) and (6) into Eq.(20) gives
\begin{equation}
v_\phi   =  r(J/r^2)(e^{\nu} /E).
\end{equation}
Finally, using Eq.(3) for the impact parameter $b$ in Eq.(21), we get
\begin{equation}
v_\phi   =  (b/r)e^{\nu}
\end{equation}
(Apparently, a constant orbital velocity $v_\phi$ of  any given
NZRP with the impact parameter  $b$  is  obtained  for a structure 
with
\[e^{\nu} \propto  r, \]
\vspace{0.2cm}
or,
\begin{equation}
\hspace{3.0cm} e^{\nu}  =  Cr, \,\,\,\,\,\,\,\,\,\,\,\,C = {\rm const.}
\end{equation}
\noindent
It is interesting to note that a configuration of 
this type corresponds to  the  well-known  non-terminating  exact 
solution with the density variation $\rho \propto  (1/r^2)$ and the pressure  $P   =  
(\rho /3)$, which lead to unphysical conditions in the sense  that 
both the pressure and density are infinite at the centre.)
\medskip

     One can mention that Hojman et al. [19] have  obtained  
an  expression  for the orbital  velocity, 
$v_{\phi,c}$   (say) by assuming that all test-particles move  in 
circular orbits, that is,  they  have  simply substituted  the  value  of 
$(d\phi /dt)$ under this {\em arbitrary} assumption  into  Eq.(19)
(see, e.g., Eq.(4.8.25) on p. 188 of Ref. [31]), and  obtained 
the orbital velocity
\begin{equation}
v_{\phi, c}   =  {[(r/2)d(e^{\nu})/dr]}^{1/2}.
\end{equation}
Using the TOV [32, 33] equation, Eq.(24) can be written as
\begin{equation}
v_{\phi, c}   =  (e^{(\nu+\lambda)/2} /\sqrt{2}) (8\pi Pr^2 + 1 - e^{-\lambda})^{1/2}.
\end{equation}
Eq.(25) indicates  that  for  any  regular  (in  the  sense 
discussed in [34])  density  distribution, one
always  obtains a rising  rotation   curve   (since both
$ e^{\nu} $ and $e^{\lambda}$ are increasing functions of $r$ in 
any regular solution). The slope of the curves are determined 
by the density distribution considered to specify the galaxy.  One 
may  choose  the  density  variation  required  according  to  the 
problem. For example, one may choose between fastest variation  of 
density $(\propto 1/r^2)$ to obtain almost constant  rotation curves and the 
smoothest density variation (i.e., the constant density  solution) 
to obtain the steepest rising rotation curve. However,  such  models 
are unable to give declining rotation curves. Furthermore, the 
total amount of dark matter in such  models  is  not  certain. 
Moreover, if one wishes  to  obtain  almost  flat  rotation
curves by using the $1/r^2$ density distribution (as used in [19]), 
it makes  the  model  artificial since, to avoid the central  singularity of the
$(1/r^2 )$ density distribution, one has to replace the central core of the model with some other, nonsingular density distribution.
\medskip

We can write Eq.(22) in terms of the variables $\beta$ and $y$ [Eqs.(14) and (15)] as
\begin{equation}
v_{\phi}  =  (\beta/y)e^{\nu}.
\end{equation}
The mean stellar velocity in various galaxies corresponds to a  value of
$\sim 300 $ km\, sec$^{-1}$ [20], which gives the 
specific energy of a star $E \cong  1.0000$.  If  we  substitute  the 
typical value of the  orbital  velocity $\sim 150  -  250 $ km\,  sec$^{-1}$,
measured in various spiral galaxies (see, e.g., Fig. 2 of [1]) into
Eq.(26), we obtain the typical value  of  the 
compaction parameter $u$ of these galaxies as $u \sim (1.270 - 2.115) \times 10^{-4}$ 
[e.g., substituting $(\beta /y)$
from Eq.(17) into Eq.(26) for $\psi = \psi_i > \psi_0$ and $E = 1$, as shown
in Table 3, and using the boundary condition that at $r = a$ the
internal solution matches with the exterior Schwarzschild solution, that is,
at $y = 1, \,e^\nu = e^{-\lambda} = (1 - 2u)$]. This is  about 127-212  times 
higher than the typical value of the compaction parameter, $ u \sim  10^{-6}$, 
presently believed for these galaxies.
\medskip

     Notice that for the value of $u (\sim 10^{-4})$ obtained  here,  the
metric function $e^{\nu}$ remains almost constant $(\cong 0.9999)$ from centre to the  outer  region 
of the configuration, and  
Eq.(26) always corresponds to  a  declining  rotation
curve for {\em a given NZRP}. However, as we have discussed  in detail in Sec.1, 
for a typical spiral galaxy (size $\sim 10^4$ pc), 
the time scale required to measure the deviation  in  the  orbital 
velocity for a {\em particular}  NZRP is
of the order of megayears, which is meaningless
in  practice  (even  if  the  measurement  could   be   possible). 
Therefore, the only remaining physically viable option (which also seems  to 
be likely) to  explain  various  observations  of  flat,  slightly 
rising, and even declining orbital velocities in  spiral  galaxies 
is to assume that we are  observing  the  {\em loci} of 
various trapped NZRPs, originated simultaneously from various  points 
inside the structure from central  to  the  outer  region  of  the 
galaxy. According to  Eq.(26), the initial orbital  velocity 
of a NZRP, originating from a specific point $(y_i)$, depends mainly on 
$b$ (which is a constant of motion and  depends  only on the {\em initial 
conditions}, see. e. g. [28]). Thus  we can construct  various  models  of  spiral 
galaxies,  corresponding  to  flat,  slightly  rising,  and   even 
declining (initial) orbital velocities of its constituents (NZRPs), by reconstruction
of the {\em initial conditions} of the NZRPs. The {\em loci} of  these  NZRPs
give specific  shapes  of  the  spirals,  which  may  be 
obtained according to the choice of the {\em initial conditions}. A possible explanation
of the observational fact that the majority of spiral galaxies show almost flat or slightly
rising rotation curves is given in the next section.

\vspace {0.5cm} 

\noindent{\large \bf 4.\,\,\,Discussion:  constant,  slightly rising, and declining
rotation curves of spiral galaxies. The DMC model and the value of $\Omega$}

\vspace {0.5cm}
\noindent
     To discuss the proposed models, we  first  work  out
the term $(\ss /y)$ appearing in Eq.(18)  and  then  find  the 
trapping angles $\psi_0$ for various NZRPs $(E \cong 1.0000)$ emitted  from 
various points $(y = y_i)$ inside a static and physically  realistic  mass 
distribution  described  by  Tolman's  type  VII   solution   with 
vanishing surface density [32, 35, 36, 22]. 
Table 1 shows various $\ss /y_i$ values at different points $(y = y_i)$  within 
the structure. Using Table 1, we  have  obtained  the  trapping 
angles $\psi_0$ as shown in Table 2 for various $y$  values  inside  the 
configuration corresponding to  some  typical  values  of $ u $.  The 
solution is given in a simple and convenient form in [35, 36, 22]. 
[Note that one may use some  other physically realistic equation  of  state  or  exact 
solution for DMC, but the  compaction  parameter  of  the  whole 
configuration will  remain  the  same,   keeping   the   results 
unaffected. However, in those cases the calculations for obtaining 
the trapping angles would be rather complicated.]
\medskip

     Having obtained the trapping angles $\psi_0$, we can work  out
constant, slightly rising and even declining instantaneous orbital 
velocities corresponding to various trapped NZRPs (with $\psi \ > \psi_0$) 
emitted simultaneously from near-central to  outer  region 
of  the  configuration,  such  that  their  {\em loci}   comprise a 
spiral-like curve with a specific shape according to  the  initial 
conditions. Table 3 describes various {\em initial  conditions}  imposed 
on NZRPs emitted from the near-central region  (e.g., $ y  = 
0.2$) to the surface ($y = 1.0$) of  the  configuration.  Their 
initial  orbital  velocities  are (i)  almost  constant,  (ii) 
slightly rising, and (iii) even declining throughout  the  configuration ranging from  100 to 339 km\, sec$^{-1}$,
 corresponding to configurations  with  compaction  parameters
$1.270 \times 10^{-4}, 1.710 \times 10^{-4}$, and $2.115 \times 10^{-4}$, respectively.
\medskip

     For illustration, we have drawn the {\em loci}  of  NZRPs  for 
the initial conditions corresponding to the first three columns of Table  3.  The 
loci indicated by the points A, B, C, D, E form a 
spiral-like curve and correspond to an almost constant orbital  velocity, 
$v_\phi \cong 203\, $km\, sec$^{-1}$ [Fig.2].  The  compaction  parameter  of  the 
structure turns out to be $\sim  1.71 \times 10^{-4}$. Similarly, as  shown  in 
Fig.3, the NZRPs indicated by the points A, B, C, 
D, E, are emitted from the  points, $ y $  =  1.0, 
0.8, 0.6, 0.4, and 0.2 inside the  configuration  with the compaction 
parameter, $u \sim 2.12  \times  10^{-4}$.  They  comprise  a  slightly  rising 
rotation curve, with orbital velocities ranging from 200 to 251 km\, sec$^{-1}$.
The locus of the NZRPs marked with  A,  B,  C,  D,  E in Fig.4
corresponds to a declining rotation curve .  The  orbital 
velocity decreases from point E ($\sim$ 203 km\,  sec$^{-1}$)  to  
point A  ($\sim$ 150  km\,  sec$^{-1}$)  inside a configuration  with 
the compaction parameter $u \sim 1.27 \times 10^{-4}$.
\medskip

     The actual trajectories of these NZRPs (A, B, C, D, E) 
are also obtained using Eqs.(16) and (17) for  the  metric 
parameters $e^{\nu}$  and $e^{\lambda}$ given by Tolman's type VII solution with
vanishing surface density [35, 36, 22] and  shown in the 
respective figures (2-4) with dashed  lines.  In  all 
trajectories a test-particle moves inward  until  it  reaches  a 
minimum distance $(y_{min})$ given by the equation [28, 35]
\begin{equation}
y_{min}^2 [(e^{-\nu})_{min}  -  E^{-2}]  -  \beta^2   =   0,
\end{equation}
\noindent
where  $ (e^{-\nu})_{min} $ is  the  value  of $ e^{-\nu} $ at $ y = y_{min} $, and   
$\beta  =  y_i ((e^{-\nu})_i - E^{-2})^{1/2} \sin \psi_i  $.
Then it remains  confined  in  a  circular  orbit  at the  minimum
distance $(y_{min})$.
\medskip

     Table 4 indicates various values of  this  minimum  distance 
$(y_{min})$ from the centre of each configuration considered. 
The small  circles  in  Figs.2-4  indicate  the 
boundary of the central region where all  NZRPs
considered here are trapped in circular orbits of the radius 
$y_{min}$. The final orbital velocity   at  $y = y_{min}$
is from 6544 to 8445 km\, sec$^{-1}$ [note 
that this velocity range also corresponds  to  various  NZRPs 
trapped within $y = 0.05$ (as shown in  Figs. 
2-4), even if they are emitted from inside this circle,  because 
they are always trapped  after  following  the  corresponding 
trajectories ending with a circular orbit given by Eq.(27).]. 
The values of escape velocities at
the minimum distance $y_{min}$ [denoted $v_{esc} (y_{min})$] and
at the surface of the structure [denoted $v_{esc} (a)$] are also
shown in Table 4. It is apparent that the values of escape velocities
are always higher than the particle velocities at the
respective points. The findings of this paper   might  be taken as a 
successful alternative to the present assumption on the 
presence of SBHs at  the  centres  of  various 
galaxies (made to explain various observations of  high 
gas velocities near galactic centres). However, if SBHs are certainly
present at the centres of some or many galaxies, the present scenario can provide a clue for
explaining their formation (see also [35, 22]).
\medskip

Now, we turn our attention to the observational evidence that most of the 
spirals show almost flat or slightly rising rotation curves and its possible 
explanation based on the present model. Consider, e.g., the trapping angle
$\psi_0$ at the surface of the configuration ($y = 1.0$) with the compaction parameter
$u = 1.71 \times 10^{-4}$, corresponding to the value $177^{\circ}.882$ [Table 2]. The NZRP emitted from
any point inside the configuration at an angle $\psi_i > 177^{\circ}.882$ will be 
gravitationally trapped. As shown in Table 3, NZRPs emitted from any point inside
the structure ($u = 1.71 \times 10^{-4}$) at an angle $\psi_i = 177^{\circ}.906$, constitute a declining rotational
curve from the central region to the surface, whereas those emitted
at an angle $\psi_i > 177^{\circ}.906$ constitute a slightly rising or almost flat 
rotation curve. Evidently, the possibility of NZRP emission at an angle 
$\psi_i > 177^{\circ}.906$ is always higher than the same near $\psi_i = 177^{\circ}.906$,
probably explaining the reason of why the majority of spiral galaxies show almost flat or
slightly rising rotation curves.
\medskip

     Furthermore, we  can  estimate
the density parameter $\Omega [\equiv (\rho_0 /\rho_{crit})$, where $\rho_0$ is the present  mass 
density of  the  Universe] in  the  following 
manner. It is well known that the population of  the  Universe  is 
dominated by spiral galaxies. The density parameter $\Omega$ for  these 
galaxies alone (including dark matter of 3 to 5 times the luminous matter density
in a visible galaxy, with a typical size
$\sim 10$ kpc) is estimated as $\sim   0.002 - 0.003$ [1, 23, 24]. 
According to the present study, we estimate that a typical galaxy contains a dark matter mass about 
127 to 212 as much as luminous matter. If we assume that our description  applies to all spiral galaxies and, 
as was mentioned  above,
that the Universe is  dominate  by  these  galaxies,  the  density 
parameter of the Universe will be at most about 0.127-0.212, leading to an open model  of  the  Universe  
with  an  age 
estimate around 16.8-17.6 Gyr (for the Hubble constant $H = 50\,
$ km\, sec$^{-1}$\, Mpc$^{-1}$ as regarded by Rubin [1]  for  observations  of 
spiral galaxies). This is significantly higher than 
the globular clusters age $\sim  13 - 15 $ Gyr [25, 26].

\vspace {0.5cm} 

{\large \bf \noindent Acknowledgment}

\vspace {0.5cm} 

\noindent
The author acknowledges  State  Observatory, 
Nainital for providing library and computer-centre facilities.

\medskip
\newpage

{\large \bf \noindent Appendix}

\vspace {0.5cm} 

\noindent
The compaction  parameter  $u$ is  a  dimensionless 
parameter which represents the total mass to size  (radius)  ratio 
of a static spherical configuration. Thus it is a measure  of 
its compactness, saying how much  mass is 
accumulated in a sphere of radius $a$ so that the configuration 
does not lose its hydrostatic equilibrium.  The  parameter 
is defined as
\vspace {0.15cm} 

\begin{equation}
u  =  GM/ac^2,                                               
\end{equation}
\vspace {0.1cm} 

\noindent where  $ M $ is  the  total  mass  and $ a $ is  the  radius   of   the 
configuration; $ c $ is the speed of light  in  vacuum, 
and $ G $ is the Newtonian gravitational constant.
\vspace {0.15cm} 

        GR is 
essentially a theory of gravity  in  which  the  gravitational 
force is regarded as an in built property of the space-time  curvature 
produced by a mass distribution. Therefore, it  is  convenient  to 
measure all physical quantities  (like  mass,  density,  pressure, 
etc.) in units of length. This can be done by choosing
\vspace {0.15cm} 

\begin{equation}
G  =  c  =  1,                                                 
\end{equation}
\vspace {0.1cm} 

\noindent so that we obtain various quantities in ``geometrized units'' as: 
$u  \equiv  (M/a)$,
$1\, {\rm gm}  \equiv  0.742 \times 10^{-28} \, {\rm cm}$,
$1\, {\rm sec}  \equiv  3 \times 10^{10} \, {\rm cm}$, etc. Thus, in particular
\vspace {0.15cm}

\begin{equation}
1 M_{\odot}  =  2 \times 10^{33} \,{\rm gm}  \equiv 1.484 \, {\rm km} \equiv 5 \times 10^{-16} \, {\rm kpc}.     
\end{equation}

\vspace {0.3cm} 

\noindent
It  is  apparent  from  Eq.(28) that the  ``compaction  parameter''  is 
equivalent to the ``dimensionless  gravitational  potential'' (i.e., the gravitational potential 
in units of $c^2$), which is the 
``dimensionless  gravitational  energy  per  unit  mass''. 
However, an exact expression for  the  gravitational  energy  per 
unit mass in GR, denoted by $\alpha_p$, in terms of  $u$
is given by [20]

\begin{equation}
\alpha_p  =  (u_p/u) - 1, \,\,\,\,\,\,\, u_p  =  (M_p/a),                                     
\end{equation}

\vspace {0.25cm} 

\noindent where $a$ is the radius and $ M_p $ is the proper mass  of  the 
configuration given by
\vspace {0.2cm} 
\begin{equation}
M_p  =  \int_{0}^{a} 4 \pi \rho r^2 e^{\lambda/2} dr
\end{equation}
\vspace {0.15cm} 
                                                              
\noindent Here $ \rho $ is the  energy-density and 
$e^{\lambda}$ is the metric coefficient such that
\vspace {0.15cm} 

\begin{equation}
e^{-\lambda}  =  1  -  [2m(r)/r],
\end{equation}
\vspace {0.15cm} 

\noindent $ m(r) $ being the mass contained inside the radius $r$.

\pagebreak
\baselineskip 1.0cm
{\large \bf References}

\begin{itemize}
\item[{1.}]
  V.C. Rubin, {\it in: \/} ``Bright Galaxies, Dark Matters'',
        AIP Press, American Institute of Physics, 1997.
\item[{2.}]

   C. Carignan and D. Puche, {\it Astron. J. \/} {\bf 100}, 394 (1990).

\item[{3.}]

   S. Casertano and J.H. van Gorkom, {\it Astron. J. \/} {\bf 101}, 1231 (1991).
 
\item[{4.}]

   T.S. van Albada and R. Sancisi, {\it Phil. Trans. R. Soc.
          London A \/} {\bf 320}, 447 (1986).

 \item[{5.}]

      R. Sancisi and T.S. van Albada, {\it in: \/} ``Dark Matter in the
        Universe'', ed. J. Kormendy and G. R. Knapp, IAU Symposium No. 117,  
        Reidel, Dordrecht, 1987.

 \item[{6.}]
 
     M. Milgrom, {\it Astrophys. J. \/} {\bf 270}, 365 (1983).
           	
 \item[{7.}] 
 
  K.G. Begelman, A.H. Broeils and R.H. Sanders, {\it Mon. Not. R. Astron. Soc. \/} 
  {\bf 249}, 523 (1991).

 \item[{8.}]

  J. Kormendy and D. Richstone, {\it Ann. Rev. Astron. Astrophys. \/} {\bf 33}, 581 (1995).

 \item[{9.}]
 
     E. Maoz, {\it Astrophys. J. \/} {\bf 447}, L91 (1995).
 
 \item[{10.}]
 
     J. Kormendy et al., {\it Astrophys. J. \/} {\bf 459}, L57 (1996).
 
  \item[{11.}]     

     F.C. van den Bosch, W. Jaffe, {\it in: \/} ``The  Nature  of 
          Elliptical Galaxies'', ed. M. Arnaboldi, G. S. Da Costa and P. Saha,
          ASP, San Francisco, 1997.

  \item[{12.}]     

     J. Kormendy et al., {\it Astrophys. J. \/} {\bf 482}, L139 (1997).

  \item[{13.}]     

     R.P. van der Marel, N. Cretton, T. de Zeeuw and H.-W. Rix, {\it Astrophys. J. \/}
          {\bf 493}, 613 (1998).
   
  \item[{14.}]    
  
     A. Marconi et al., {\it Mon. Not. R. Astron. Soc. \/}, in press, (1998).
     
 \item[{15.}]    
  
     E. Maoz, {\it Astrophys. J. \/} {\bf 494}, L181 (1998).
     
\item[{16.}] 

    G.C. Stewart, C.R. Canizares, A.C. Fabian and P.E.J. Nulsen, {\it Astrophys. J. \/}
        {\bf 278}, 536 (1984).
  
\item[{17.}]      

    I. Hernquist, {\it Astrophys. J. \/} {\bf 356}, 359 (1990).
         
\item[{18.}]      

   E. Maoz and J.D. Beckenstein, {\it Astrophys. J. \/} {\bf 353}, 59 (1990).
   
\item[{19.}]        
  
     R. Hojman, L. Pena and N. Zamorano,  {\it Astrophys. J. \/} {\bf 411}, 541 (1993).

\item[{20.}]         

     Ya.B. Zeldovich and I.D. Novikov, {\it in: \/} ``Relativistic Astrophysics", 
     Vol.1, Chicago University Press, Chicago, 1978.

\item[{21.}]
 
      Y. Sofue et al., {\it Astrophys. J. \/} {\bf 523}, 136 (1999).      
     
\item[{22.}]      

      P.S. Negi and M.C. Durgapal, {\it Astrophys. \& Space Science \/} {\bf 275}, 185 (2001).
        
 \item[{23.}]     
     
     K. Freeman, {\it in: \/} ``Les Houches Summer School on Cosmology'', ed. R.
     Schaeffer, 1994.
 
 \item[{24.}]

     M. Rowan-Robinson, {\it in: \/} ``Les Houches Summer School on
     Cosmology'', ed. R. Schaeffer, 1994.

\item[{25.}] 

     D.A. van den Berg, M. Bolte and P.B. Stetson, {\it Ann. Rev. Astron. Astrophys. \/}
          {\bf 34}, 461 (1996).

\item[{26.}]        

     G.A. Tammann, {\it in: \/} ``Relativistic Astrophysics'', ed. B. J. T. 
         Jones and D. Markovic, Cambridge University Press, Cambridge, 1997.

\item[{27.}] 

     J.L. Synge, {\it Mon. Not. R. Astron. Soc. \/} {\bf 131}, 463 (1966).

\item[{28.}] 
 
     P.S. Negi, A.K. Pande and M.C. Durgapal, {\it Astrophys. J. \/} {\bf 406}, 1 (1993).
   
\item[{29.}]      

     S. Chandrasekhar, {\it in: \/} ``The Mathematical Theory of Black Holes'',
        Oxford University Press, Clarendon, 1983.
  
\item[{30.}]          

     P.S. Negi, K. Pandey and A.K. Pande, {\it Astrophys. \& Space Science \/} {\bf 176},
       131 (1991).
 
\item[{31.}]      

  S. Weinberg, {\it in:\/} ``Gravitation and Cosmology'',

          Wiley, New York, 1972.

\item[{32.}]          

     R.C. Tolman, {\it Phys. Rev. \/} {\bf 55}, 364 (1939).
   
\item[{33.}] 

     J.R. Oppenheimer and G.M. Volkoff, {\it Phys. Rev. \/} {\bf 55}, 375 (1939).
  
\item[{34.}]  

     H.A. Buchdahl, {\it Phys. Rev. \/} {\bf 116}, 1027 (1959).
     
\item[{35.}]

     P.S. Negi and M.C. Durgapal, {\it Astrophys. \& Space Science \/} {\bf 245}, 
       97 (1996).
  
\item[{36.}]

      P.S. Negi and M.C. Durgapal, {\it Gen. Rel. Grav. \/} {\bf 31}, 13 (1999).

\end{itemize}


\pagebreak

\begin{table}
\begin{center} 
{\bf Table 1.}

\hspace{1.0cm}

\begin{tabular}{cccccc}

\hline

$ $ & $E = 1.0000$  \\

\hline

\hline

$y_i$ & $u_i$ & ${\ss /y_i}$  \\

\hline

0.2000   &   0.1630   &   3.3955  \\

0.4000   &   0.1696   &  3.1909   \\

0.6000   &   0.1827   &  2.8633   \\

0.8000   &   0.2071   &  2.4413   \\

1.0000   &   0.2500   &  2.0000   \\

\hline

\end{tabular}
\end{center}
\end{table}

\begin{table}
\begin{center}
 {\bf Table 2.}

\hspace{1.0cm}

\begin{tabular}{cccccc}

\hline

$ $ & $ $ & ${E=1.0000}$  \\

\hline

$ $ & $ $ & ${\psi_0\, {\rm for\, \ various}\,  y_i}$  \\

\hline

${u(10^{-4})}$ & 0.2000 & 0.4000 & 0.6000 & 0.8000 & 1.0000  \\

${\downarrow}$                                                \\

\hline

1.27000 & ${177^{\circ}.706}$ & 177.756 & 177.847 & 177.990 & 178.174  \\

1.71000 & 177.338 & 177.396 & 177.502 & 177.667 & 177.882  \\

2.11500 & 177.040 & 177.104 & 177.221 & 177.406 & 177.644  \\

\hline
\end{tabular}
\end{center}
\end{table}

\newpage

\begin{table}
\begin{center} 
{\bf Table 3.}

\hspace{1.0cm}

\begin{tabular}{cccccccc}

\hline

$ $ & $ $ & $ $ & ${E=1.0000}$  \\

\hline

$u(10^{-4})$ & $y_i$ & ${\psi_i}$ & $v_{\phi}(\rm km\, sec^{-1})$ & ${\psi_i}$ & $v_\phi(\rm km\, sec^{-1})$ & ${\psi_i}$ & $v_\phi(\rm km\, sec^{-1})$  \\

\hline

1.710 & 0.2000 & 178$^{\circ}.450$ & 202.8 & 178$^{\circ}.852$ & 150.2 & 177$^{\circ}.906$ & 273.9  \\

1.710 & 0.4000 & 178$^{\circ}.387$ & 202.8 & 178$^{\circ}.720$ & 160.9 & 177$^{\circ}.906$ & 263.1 \\

1.710 & 0.6000 & 178$^{\circ}.275$ & 202.8 & 178$^{\circ}.506$ & 175.6 & 177$^{\circ}.906$ & 246.1  \\

1.710 & 0.8000 & 178$^{\circ}.111$ & 202.8 & 178$^{\circ}.275$ & 185.2 & 177$^{\circ}.906$ & 224.8  \\

1.710 & 1.0000 & 177$^{\circ}.906$ & 202.8 & 177$^{\circ}.906$ & 202.8 & 177$^{\circ}.906$ & 202.8  \\

\hline
                                                
2.115 & 0.2000 & 178$^{\circ}.276$ & 250.8 & 178$^{\circ}.624$ & 200.2 & 177$^{\circ}.670$ & 339.0  \\

2.115 & 0.4000 & 178$^{\circ}.205$ & 250.9 & 178$^{\circ}.496$ & 210.2 & 177$^{\circ}.670$ & 325.6  \\

2.115 & 0.6000 & 178$^{\circ}.081$ & 250.8 & 178$^{\circ}.272$ & 225.9 & 177$^{\circ}.670$ & 304.5  \\

2.115 & 0.8000 & 177$^{\circ}.898$ & 250.9 & 178$^{\circ}.018$ & 236.6 & 177$^{\circ}.670$ & 278.1  \\

2.115 & 1.0000 & 177$^{\circ}.670$ & 250.9 & 177$^{\circ}.670$ & 250.9 & 177$^{\circ}.670$ & 250.9  \\

\hline

1.270 & 0.2000 & 178$^{\circ}.669$ & 150.1 & 179$^{\circ}.109$ & 100.5 & 178$^{\circ}.203$ & 202.6  \\

1.270 & 0.4000 & 178$^{\circ}.615$ & 150.0 & 178$^{\circ}.932$ & 115.7 & 178$^{\circ}.203$ & 194.6  \\

1.270 & 0.6000 & 178$^{\circ}.519$ & 150.0 & 178$^{\circ}.758$ & 125.8 & 178$^{\circ}.203$ & 182.0  \\

1.270 & 0.8000 & 178$^{\circ}.378$ & 150.1 & 178$^{\circ}.532$ & 135.8 & 178$^{\circ}.203$ & 166.2  \\

1.270 & 1.0000 & 178$^{\circ}.203$ & 150.0 & 178$^{\circ}.203$ & 150.0 & 178$^{\circ}.203$ & 150.0  \\

\hline
                                                
\end{tabular}
\end{center}
\end{table}

\begin{table}
\begin{center} 
{\bf Table 4.}

\hspace{1.0cm}

\begin{tabular}{ccccccccccc}

\hline

$ $ & $ $ & $ $ & ${E=1.0000}$  \\

\hline

$u(10^{-4})$ & $y_i$ & $\beta(10^{-4})$ & $y_{min}$ & $v_{\phi}(y_{min}) (\rm km\, sec^{-1}) $ & $v_{esc}\, (a) (\rm km\, sec^{-1}) $ & $v_{esc}(y_{min})\, (\rm km\, sec^{-1})$  \\

\hline

1.710 & 1.0 & 6.7594 & 0.0267 & 7592.0 & 5548.1 & 7595.3  \\

1.710 & 0.8 & 5.4080 & 0.0214 & 7592.7 & 5548.1 & 7596.0  \\

1.710 & 0.6 & 4.0563 & 0.0160 & 7593.2 & 5548.1 & 7596.7  \\

1.710 & 0.4 & 2.7035 & 0.0107 & 7593.5 & 5548.1 & 7596.7  \\

1.710 & 0.2 & 1.3523 & 0.0053 & 7593.6 & 5548.1 & 7596.7  \\

\hline

2.115 & 1.0 & 8.3645 & 0.0297 & 8442.0 & 6170.6 & 8446.1  \\

2.115 & 0.8 & 6.3101 & 0.0224 & 8444.0 & 6170.6 & 8448.0  \\

2.115 & 0.6 & 4.5187 & 0.0160 & 8444.7 & 6170.6 & 8448.6  \\

2.115 & 0.4 & 2.8039 & 0.0099 & 8444.0 & 6170.6 & 8448.6  \\

2.115 & 0.2 & 1.3353 & 0.0047 & 8444.6 & 6170.6 & 8449.2  \\
                                  
\hline

1.270 & 1.0 & 4.9999 & 0.0229 & 6543.8 & 4781.5 & 6546.1  \\

1.270 & 0.8 & 4.4333 & 0.0203 & 6543.7 & 4781.5 & 6546.1  \\

1.270 & 0.6 & 3.6412 & 0.0167 & 6544.7 & 4781.5 & 6546.9  \\

1.270 & 0.4 & 2.5956 & 0.0119 & 6544.7 & 4781.5 & 6546.9  \\

1.270 & 0.2 & 1.3509 & 0.0062 & 6545.2 & 4781.5 & 6547.7  \\

\hline
                                                
\end{tabular}
\end{center}
\end{table}


\pagebreak

\begin{figure*}[h]
\noindent
Fig.1: The azimuthal angle $\phi$ and the emittance angle $\psi$,  at  a  radial 
distance $y(\equiv r/a)$ from the centre of spherical configuration in
$\theta = (\pi/2)$ plane. $0$ and $ a $ are the
centre and radius of the spherical   configuration, respectively. A
test-particle  is initially  emitted  from  a  point  at  radial 
distance $ y = y_i$ with $\psi = \psi_i $ and $\phi = \phi_i$.  

\end{figure*}
\pagebreak

\begin{figure*}[h]
\noindent
Fig.2: The spiral-like  curve  is  the  {\em locus}  of  NZRPs  emitted 
initially from the points with the following coordinates ($y_i,\,\,\,\psi_i,\,\,\,\phi_i$): A (1.0, \,\,\,$177^{\circ}.906$, \,\,\, 
$ 0^{\circ}$), B (0.8,\,\,\, $178^{\circ}.111$,\,\,\, $10^{\circ} .0)$,
C (0.6, \,\,\,$178^{\circ} .275$, \,\,\,$30^{\circ} .0$),
D (0.4, \,\,\,$178^{\circ} .387$, \,\,\,$50^{\circ} .0$),
E (0.2, \,\,\,$178^{\circ} .450$, \,\,\,$60^{\circ} .0$),
such that their (initial) orbital velocity remain  constant,  about 
203 km\, sec$^{-1}$. The dashed curves represent the NZRP trajectories ending in circular 
orbits of minimum distances $y_{min}$ with finally reached values of the orbital
velocity $v_{\phi}(y_{min}) \sim 10^4 \,{\rm km\,\, sec}^{-1}$, as shown in Table 4. These NZRPs are trapped 
in circular orbits of  radius 
$y_{min}$ in the region shown by a small circle of radius $y = 0.05$. The compaction 
parameter is $u = 1.71 \times 10^{-4}$.

\vspace{12.0cm}
\end{figure*}
\newpage
\begin{figure*}[h]
\noindent
Fig.3: The spiral-like  curve  is  the  {\em locus}  of  NZRPs  emitted 
initially from the points with the following coordinates ($y_i,\,\,\,\psi_i,\,\,\,\phi_i$):
A (1.0, \,\,\,$177^{\circ}.670$,
\,\,\,$0^{\circ}$), B (0.8, \,\,\,$178^{\circ}.018$, \,\,\,$30^{\circ} .0$),
C (0.6, \,\,\,$178^{\circ} .272$, \,\,\,$70^{\circ} .0$),
D (0.4, \,\,\,$178^{\circ} .496$, \,\,\,$100^{\circ} .0$),
E (0.2, \,\,\,$178^{\circ} .624$, \,\,\,$150^{\circ} .0$),
such that their (initial) orbital velocities slightly rise from 200 km\, sec$^{-1}$
at point E to $\sim$ 251 km\, sec$^{-1}$ at point A. The dashed curves represent the NZRP trajectories ending
in  circular orbits of minimum distance $y_{min}$ with finally reached orbital
velocities $v_{\phi}(y_{min}) \sim 10^4 \,{\rm km\,\, sec}^{-1}$, as shown in Table 4. The  boundary 
inside which these NZRPs are trapped in circular orbits of  radius 
$y_{min}$ is shown by a small circle of radius $y = 0.05$. The compaction parameter is
$u = 2.115 \times10^{-4}$.
\vspace{12.0cm}
\end{figure*}

\newpage
\begin{figure*}[h]
\noindent
Fig.4: The spiral-like  curve  is  the  {\em locus}  of  NZRPs  emitted 
initially from the points with the following coordinates ($y_i,\,\,\,\psi_i,\,\,\,\phi_i$): A (1.0, \,\,\,$178^{\circ}.203$, 
\,\,\,$0^{\circ}$), B (0.8, \,\,\,$178^{\circ}.203$, \,\,\,$20^{\circ} .0$),
C (0.6, \,\,\,$178^{\circ}.203$, \,\,\,$40^{\circ} .0$),
D (0.4, \,\,\,$178^{\circ}.203$, \,\,\,$70^{\circ} .0$),
E (0.2, \,\,\,$178^{\circ}.203$, \,\,\, $110^{\circ} .0$),
such that their (initial) orbital velocities decline from 202.6 km\, sec$^{-1}$
at point E to  about 150 km\, sec$^{-1}$ at point A. The dashed curves
represent the NZRP trajectories ending in  circular 
orbits of minimum distance $y_{min}$ with finally reached orbital
velocities $v_{\phi}(y_{min}) \sim 10^4 \,{\rm km\,\, sec}^{-1}$, as shown in Table 4. The  boundary 
inside which these NZRPs are trapped in circular orbits of  radius 
$y_{min}$  is shown by a small circle of radius $y = 0.05$. The compaction 
parameter is $u = 1.270 \times 10^{-4}$.
\vspace{12.0cm}
\end{figure*}

\end{document}